\documentclass[a4paper,12pt,openany]{article}
\usepackage[top=20truemm,bottom=20truemm,left=23truemm,right=23truemm]{geometry}
\usepackage[]{hyperref}
\usepackage{amsmath}
\usepackage{ascmac}
\usepackage{amssymb}
\usepackage[]{graphicx}
\graphicspath{{./figs/}}

\title{D5-brane in AdS black holes with nonzero gauge flux}
\author{Koichi Nagasaki}
\date{\today}

\begin{document}
\vspace{1cm}

\begin{center}
	{\LARGE D5-brane in AdS black holes with nonzero gauge flux}\\
\vspace{2cm}
	{\large Koichi Nagasaki}\footnote{koichi.nagasaki24@gmail.com}\\
\vspace{1cm}
	{\small Department of Physics, 
	Toho University,\\
	Address: 2 Chome-2-1 Miyama, Funabashi, Chiba 274-8510, Japan}
\end{center}
\vspace{1.5cm}

\abstract{
We find the probe D5-brane solution on the black home spacetime which is asymptomatically $\mathrm{AdS}_5\times S^5$.
These black holes have spherical, hyperbolic and toroidal structures.
Depending on the gauge flux on the D5-brane, the D5-brane behaves differently. 
By adding the fundamental string, the potential energy of the interface solution and the Wilson loop is given in the case of non zero gauge flux. 
}

\vspace{1cm}
\tableofcontents
\vspace{2cm}
\newpage
\section{Introduction}
In order to reveal the properties of the superstring theory, the AdS/CFT correspondence \cite{Maldacena:1997re,Witten:1998qj,Aharony:1999ti} have been studied in many works.
It allows us to use weak-coupling perturbative methods in one theory to study the other side of dual theory. 
For example, by such a holographic principle, the entanglement entropy is dual to the minimal surface on the bulk spacetime \cite{Hubeny:2007xt,Takayanagi:2012kg}.

Nonlocal operators are interesting objects to study this correspondence.
These are operators which extend the spacetime in conformal field theory.
For example, a Wilson loop is one dimensional case, two-dimensional object called a surface operator is studied in \cite{Koh:2008kt} and a codimension two object is studied in \cite{Gutperle:2020gez}.
The interface CFT \cite{Karch:2000gx,DeWolfe:2001pq,Bachas:2001vj,Gaiotto:2012np,Bak:2013uaa,Gutperle:2015hcv,Chiodaroli:2016jod,Rastelli:2017ecj,Dey:2020jlc,Herzog:2020wlo,Bachas:2020yxv} is a theory with one of such nonlocal objects which we would like to focus on in this paper.
By introducing a Wilson loop operator, the effect from the interface is found in \cite{Nagasaki:2011ue} from the corresponding Nanbu-Goto action.
This result is generalized to black hole spacetime in \cite{Nagasaki:2018dyz}.
However it restricted to the case where there is no gauge flux on the probe D5-brane.
On the boundary CFT this probe D5-brane realizes the interface and the gauge flux gives the difference of gauge groups between the both sides of it.
As we discussed in the last of \cite{Nagasaki:2020rbb}, the gauge flux dependence of the interface and gauge theory behaves differently depending on the structure of spacetime.
So we are interested in finding the behavior for non-trivial gauge flux.

On the other hand, black holes are important object to understand a quantum theory of gravity.
We treat the probe brane solution on three different black hole spacetimes which are distinguished by its curvature.
These solutions are distinguished by the horizon geometry of a $(d-2)$-dimensional Einstein manifold which has positive, zero, or negative curvature.
These have the structure of the sphere, torus and hyperbolic space.
These black holes are known as topological black holes \cite{Lemos:1995cm,Aminneborg:1996iz,Banados:1997df,Mann:1997iz,Vanzo:1997gw}.
For example, the planer case ($k=0$) is studied in detail in \cite{Lemos:1994fn,Lemos:1994xp,Cai:1996eg}
and thermodynamics of these black holes is studied in \cite{Brill:1997mf}.
Such a spacetime has the asymptotic AdS structure \cite{Birmingham:1998nr,10.1093/ptep/pty017} and then to be a very interesting example for testing the AdS/CFT dualities.
We put the probe D5-brane on these black hole spacetimes.
Then, it realizes the interface on the AdS boundary.
We find the solution of a probe D5-brane in these black hole spacetimes and find the difference of these interface solutions.

For the AdS/CFT duality the gauge calculation and the gravity one are compared.
The gauge theory gives a good approximation on the range $\lambda\ll1$ (small 't-Hooft coupling) while the gravity theory is good for $\lambda\gg1$. 
In \cite{Nagasaki:2011ue} we compare the two sides of theory by expanding the result by the power series of $\lambda/k^2$ and the results from both sides agree in the first order. 
In this comparing the parameter $k$, which corresponds to the gauge flux on the D5-brane, has a role to expand the gauge and gravity theory results.
We give a calculation of potential energy between the interface and a Wilson loop operator. 
In the previous work \cite{Nagasaki:2020rbb} we only consider the cases where the gauge flux is zero. 
As mentioned above this gauge flux has an important role to compare the gauge and gravity sides.
Then the study of the case of non trivial gauge flux will contribute the understanding the AdS/CFT duality. 

The construction of this paper is as follows: 
In Section \ref{sec:BHsolutions} we review the black hole solution and give the equation of motion for the probe D5-brane. 
By studying this equation of motion, we find the difference between three different topologies --- for black holes which has hyperbolic structure there is no solution which extends to the center of AdS spacetime, namely, the black hole horizon.
In Section \ref{sec:Wilson_loop} we introduce a fundamental string which extends from the AdS boundary to the black hole horizon.
This corresponds to the insertion of a Wilson loop on the boundary CFT.
Finally, in Section \ref{sec:nontrivial_gaugeflux} we consider the case where the probe D5-brane have non zero gauge flux.

\section{Black hole solution}\label{sec:BHsolutions}
The solutions of the Einstein equation which are the asymptotically AdS is given as follows: 
\begin{equation}
ds_{k,d+1}^2 = -f_k(r)dt^2 + \frac{dr^2}{f_k(r)} + r^2d\Sigma_{k,d-1}^2,
\end{equation}
where $k$ distinguishes three different topologies: $k=\pm1,0$.
For $k=1$ case this solution has the structure of sphere, for $k=0$ this solution has the structure of torus and for $k=-1$ this solution has the structure of hyperbolic space. 
In the above metric the angular part is
\begin{equation}
d\Sigma_{k,d-1}^2
= \begin{cases}
d\theta^2 + \sin^2\theta d\Omega_{d-2}^2, & (k=+1)\\
d\theta^2 + d\Omega_{d-2}^2, & (k=0)\\
d\theta^2 + \sinh^2\theta d\Omega_{d-2}^2, & (k=-1)
\end{cases}.
\end{equation}
The function $f_k(r)$ is 
\begin{equation}
f_k(r) = k + r^2 - \frac{r_\mathrm{m}^{d-2}}{r^{d-2}},\;\;
r_\mathrm{m} = \frac{16\pi G_{d+1}M}{(d-1)\Sigma_{k,d-1}},
\end{equation}
where $\Sigma_{k,d-1}$ is the dimensionless volume for each geometry \cite{Chapman:2016hwi}.
Especially, in the case of sphere, it is the volume of $(d-1)$-dimensional sphere: 
$\Omega_{k-1} = 2\pi^{d/2}/\Gamma(d/2)$.
For toroidal and hyperbolic cases, these are defined by appropriately regularizing.
We define a function $s_k(\theta)$ such that 
\begin{equation}
s_k(\theta) = \begin{cases}
\sin^2\theta & (k=+1)\\  1& (k=0)\\ \sinh^2\theta & (k=-1)
\end{cases}.
\end{equation}
By this notation, we write the angular part as 
$d\Sigma_{k,d-1}^2 = d\theta^2 + s_k(\theta)d\Omega_{d-2}^2$ in the following equations.

We consider the black hole solution in $\mathrm{AdS}_5\times S^5$ where the black hole is locate in the center of the AdS spacetime.
We set the coordinates 
$(t,r,\theta,\phi,\psi)$ on $\mathrm{AdS}_5$ and $(\varphi,\chi,\phi_1,\phi_2,\phi_3)$ on $S^5$.
It is convenient to redefine the radial coordinate $y=1/r$ where $y=0$ corresponds to the AdS boundary. 
In these coordinates, the metric is 
\begin{equation}
ds_{k,5}^2 
= \frac1{y^2}\Big(-F_k(y)dt^2 + \frac{dy^2}{F_k(y)} + d\Sigma_{k,3}^2\Big),
\end{equation}
where a new metric function is
\begin{equation}
F_k(y) 
:= y^2f_k(1/y) 
= 1 + ky^2 - r_\mathrm{m}^{d-2}y^d.
\end{equation}
It goes to unity when $y\rightarrow0$. 
Then it recovers the flat AdS spacetime sufficiently far from black holes.

We introduce a probe D5-brane which extends to the directions
$t$, $\phi$, $\psi$ in $\mathrm{AdS}_5$ and $\varphi$ and $\chi$ in $S^5$ planarly and occupies one-dimensional subspace in $(r,\theta)$ plane:
\begin{equation}
y = y(\sigma),\;\;
\theta = \theta(\sigma).
\end{equation}
There exists a gauge flux on the D5-brane.
We assume it extends on the $S^2$ subspace on $S^5$.
Then our assumption is 
\begin{equation}
\mathcal F = -\kappa\; d\text{vol}[S^2]
= -\kappa\sin\varphi d\varphi\wedge d\chi.
\end{equation}

\paragraph{D5-brane action and EOM}
The D5-brane action is 
\begin{equation}
S = S_\mathrm{DBI} + S_\mathrm{WZ} 
= -T_5\int\sqrt{-\det(g_k+\mathcal F)_\mathrm{ind}} + T_5\int\mathcal F\wedge C_4.
\end{equation}
$C_4$ is the Ramond-Ramond 4-form which satisfies
$dC_4 = 4d\mathrm{vol[AdS]} + 4d\mathrm{vol}[S^5]$.
The volume form of the AdS spacetime is
\begin{align*}
d\text{vol[AdS]} = -\frac1{y^5}s_k(\theta)\sin\phi dtdyd\theta d\phi d\psi.
\end{align*}
By definition we find the $C_4$ as
\begin{equation}
C_4 = -\frac1{y^4}s_k(\theta)\sin\phi dtd\theta d\phi d\psi + 4\alpha_4,
\end{equation}
where $\alpha_4$ is the 4-form on the $S^5$ which satisfies $d\alpha_4 = d\text{vol}[S^5]$.
Taking the product $\mathcal F\wedge C_4$, $\alpha_4$ part vanishes and gives
\begin{equation}
\mathcal F\wedge C_4 
= \frac{\kappa}{y^4}s_k(\theta)\sin\phi\sin\varphi dtd\theta d\phi d\psi d\varphi d\chi.
\end{equation}
To find the DBI action we calculate the determinant
\begin{equation}
\sqrt{-\det(g_k+\mathcal F)_\text{ind}}
= \frac1{y^4}\sqrt{y'^2 + \theta'^2F_k}\; s_k\sin\phi\sqrt{1+\kappa^2}\sin\varphi. 
\end{equation}

The D5-brane action is obtained as the sum of the above terms:
\begin{align}
S &= -T_5\int_\text{AdS}dtd\sigma d\phi d\psi\int_{S^2}d\varphi d\chi
 \frac{s_k\sin\phi\sin\varphi}{y^4}
 \Big(\sqrt{1+\kappa^2}\sqrt{y'^2 + \theta'^2F_k} - \kappa\theta'\Big)\nonumber\\
&= -(4\pi)^2T_5\Delta t\;\int d\sigma
 \frac{s_k}{y^4}\Big(\sqrt{1+\kappa^2}\sqrt{y'^2 + \theta'^2F_k} - \kappa\theta'\Big).
\end{align}
To obtain the last form we performed the integral in the angular directions 
($\phi,\psi,\varphi,\chi$), where
the metrics on two spheres, $S^2\subset\mathrm{AdS}$ and $S^2\subset S^5$, are
\begin{equation}
ds_{S^2\subset\mathrm{AdS}}^2 = d\phi^2 + \sin^2\phi d\psi^2
\quad\text{and}\quad
ds_{S^2\subset S^5}^2 = d\varphi^2 + \sin^2\varphi d\chi^2.
\end{equation} 
These coordinates take the value $\phi,\varphi\in[0,\pi),\;\psi,\chi\in[0,2\pi]$.

The Lagrangian is, eliminating the whole constant factor,
\begin{equation}
\mathcal L(\sigma) 
= \frac{s_k(\theta)}{y^4}
  \big(\sqrt{y'^2 + \theta'^2F_k} - \tilde\kappa\theta'\big),
\end{equation}
where we defined $\tilde\kappa := \kappa(1+\kappa^2)^{-1/2}$ which takes the value $|\tilde\kappa|\leq1$.
The equations of motion for $y$ and $\theta$ are
\begin{equation*}
\frac{d}{d\sigma}\frac{\partial\mathcal L}{\partial y'}
 - \frac{\partial\mathcal L}{\partial y} = 0,\;\;
\frac{d}{d\sigma}\frac{\partial\mathcal L}{\partial\theta'}
 - \frac{\partial\mathcal L}{\partial\theta} = 0.
\end{equation*}

Let us fix the gauge freedom as 
$\sqrt{y'^2 + \theta'^2F_k} = 1$.
It simplifies the equations of motion as
\begin{align}
\frac{d}{d\sigma}\Big(y'\frac{s_k}{y^4}\Big)
 + 4(1-\tilde\kappa\theta')\frac{s_k}{y^5}
 - \frac12\theta'^2\partial F_k\frac{s_k}{y^4} &= 0,\\
\frac{d}{d\sigma}\Big((\theta'F_k - \tilde\kappa)\frac{s_k}{y^4}\Big)
 - (1-\tilde\kappa\theta')\frac{\partial s_k}{y^4} &= 0.
\end{align}
By dividing by the common factor, $s_k/y^4$, these are
\begin{align}
y'' + y'\frac{d}{d\sigma}\log\frac{s_k}{y^4}
 + \frac4y(1-\tilde\kappa\theta')
 - \frac12\theta'^2\partial F_k &= 0,\\
\theta''F_k + \theta'y'\partial F_k
 + (\theta'F_k - \tilde\kappa)\frac{d}{d\sigma}\log\frac{s_k}{y^4}
 - (1 - \tilde\kappa\theta')\partial\log s_k &= 0.
\end{align}
We fixed the gauge $y'^2 + \theta'^2F_k = 1$.
Let us confirm it is preserved along the integral curve.
By differentiating by $\sigma$, 
\begin{equation}
\frac{d}{d\sigma}(y'^2 + \theta'^2F_k)
= 2y'y'' + 2\theta'\theta''F_k + \theta'^2y'\partial F_k.
\end{equation}
Substituting the equation of motion,
\begin{align*}
\frac{d}{d\sigma}(y'^2 + \theta'^2F_k)
&= 2y'\Big(-y'\frac{d}{d\sigma}\log\frac{s_k}{y^4} 
  - \frac4y(1-\tilde\kappa\theta')
  + \frac12\theta'^2\partial F_k\Big)\\
&\quad
 + 2\theta'\Big(-\theta'y'\partial F_k 
  - (\theta'F_k-\tilde\kappa)\frac{d}{d\sigma}\log\frac{s_k}{y^4}
  + (1-\tilde\kappa\theta')\partial\log s_k\Big)
 + \theta'^2y'\partial F_k\\
&= -2(y'^2 + \theta'^2F_k - \tilde\kappa\theta')\frac{d}{d\sigma}\log\frac{s_k}{y^4}
 + 2(1-\tilde\kappa\theta')\frac{d}{d\sigma}\log(y^{-4})
 + 2(1 - \tilde\kappa\theta')\frac{d}{d\sigma}\log s_k\\
&= -2(y'^2 + \theta'^2F_k - 1)\frac{d}{d\sigma}\log\frac{s_k}{y^4},
\end{align*}
where the last expression is zero if we use the gauge condition. 

Summarizing the above, we find the equations of motion:
\begin{align}\label{eq:topBH_static_eom}
y'' + y'\theta'\partial\log s_k
 + \frac4y(1-y'^2-\tilde\kappa\theta')
 - \frac12\theta'^2\partial F_k &= 0,\\
\theta''F_k + \theta'y'\partial F_k
 + (\theta'^2F_k - 1)\partial\log s_k
 - \frac4y(y'\theta'F_k - \tilde\kappa y') &= 0.
\end{align}
Here the functions are given as
\begin{equation}
\partial\log s_k
= \begin{cases}
2\cot\theta& (k=1)\\
0& (k=0)\\
2\coth\theta& (k=-1)
\end{cases},\quad
F_k = \begin{cases}
1+y^2-r_\mathrm{m}^2y^4& (k=1)\\
1-r_\mathrm{m}^2y^4& (k=0)\\
1-y^2-r_\mathrm{m}^2y^4& (k=-1)
\end{cases}.
\end{equation}

\begin{figure}[h]
	\begin{minipage}[t]{0.5\linewidth}
	\includegraphics[width=8cm]{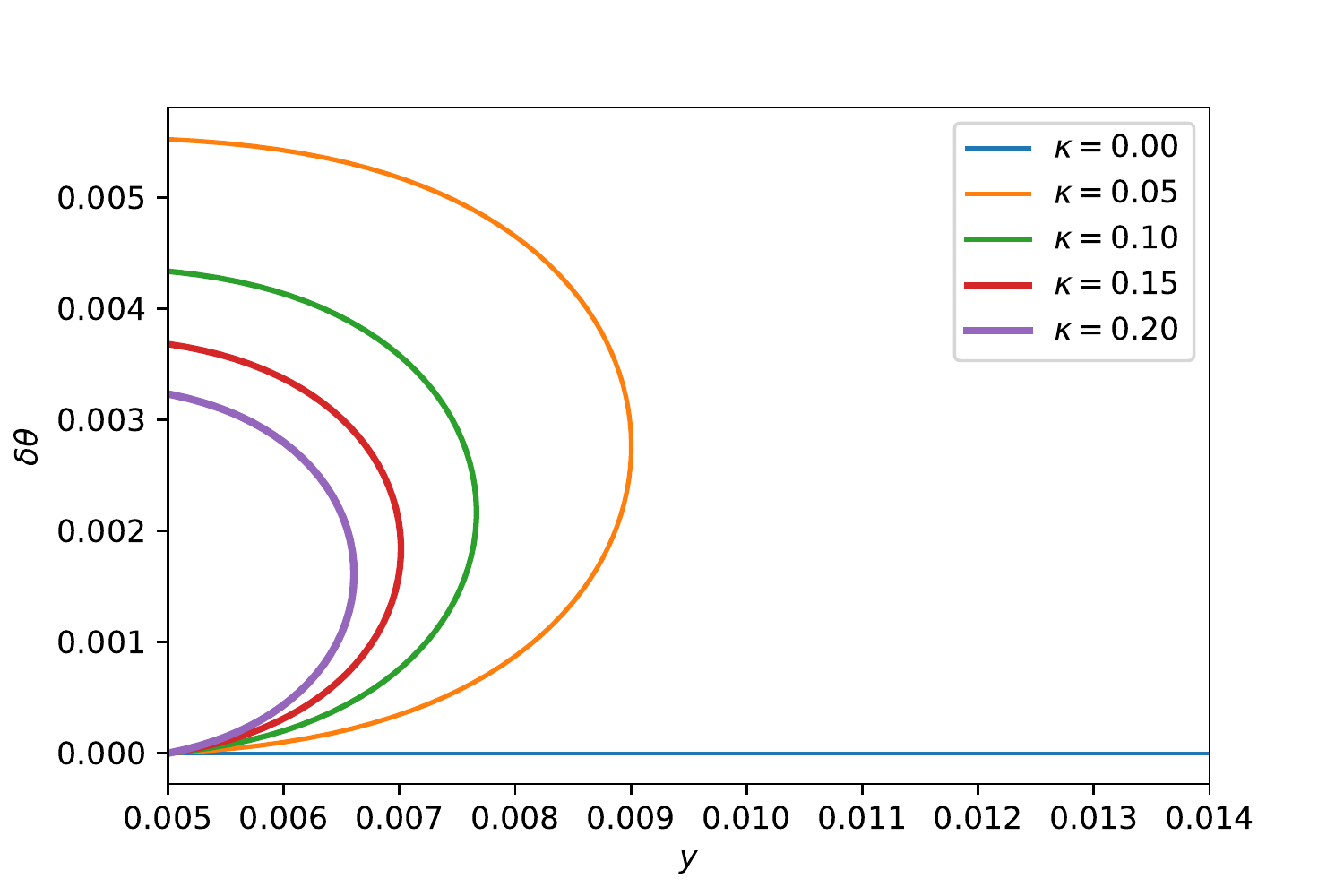}
	\caption{Gauge flux dependence (spherical case)}
	\label{fig:topBH_static_s}
	\end{minipage}
\hspace{0.01\linewidth}
	\begin{minipage}[t]{0.5\linewidth}
	\includegraphics[width=8cm]{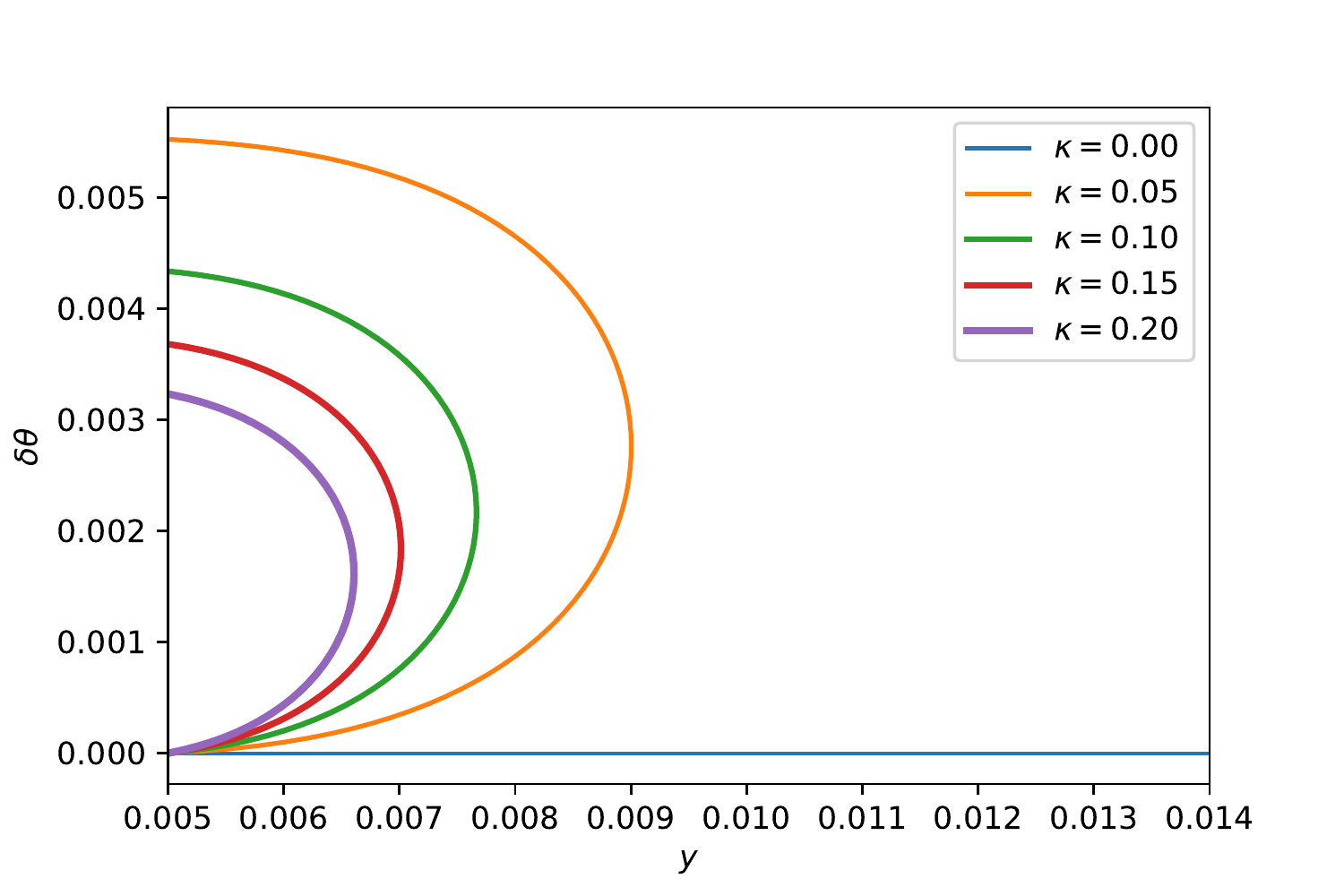}
	\caption{Gauge flux dependence (toroidal case)}
	\label{fig:topBH_static_t}
	\end{minipage}
\hspace{0.01\linewidth}
	\begin{minipage}[t]{0.5\linewidth}
	\includegraphics[width=8cm]{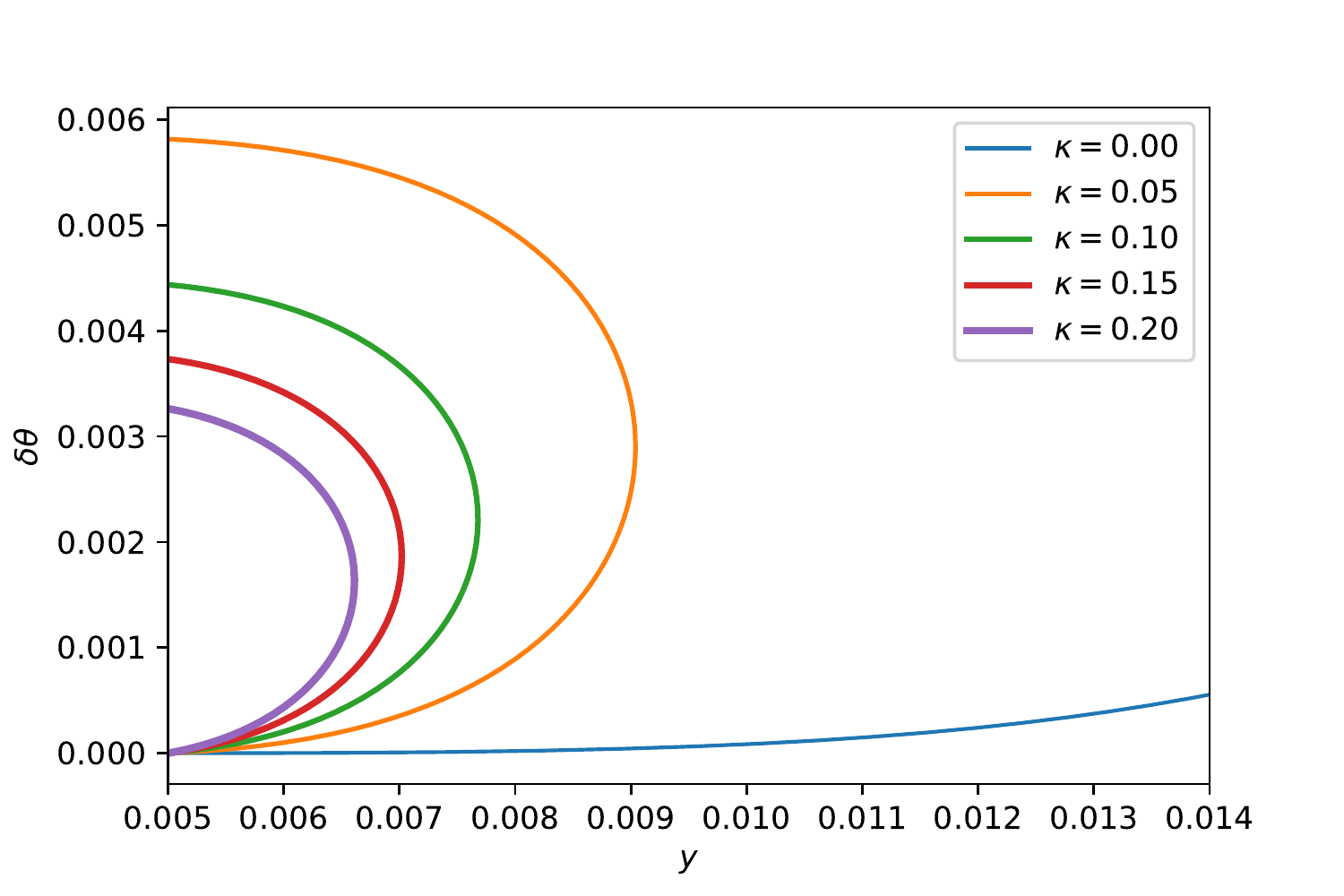}
	\caption{Gauge flux dependence (hyperbolic case)}
	\label{fig:topBH_static_h_1}
	\end{minipage}
\hspace{0.01\linewidth}
	\begin{minipage}[t]{0.5\linewidth}
	\includegraphics[width=8cm]{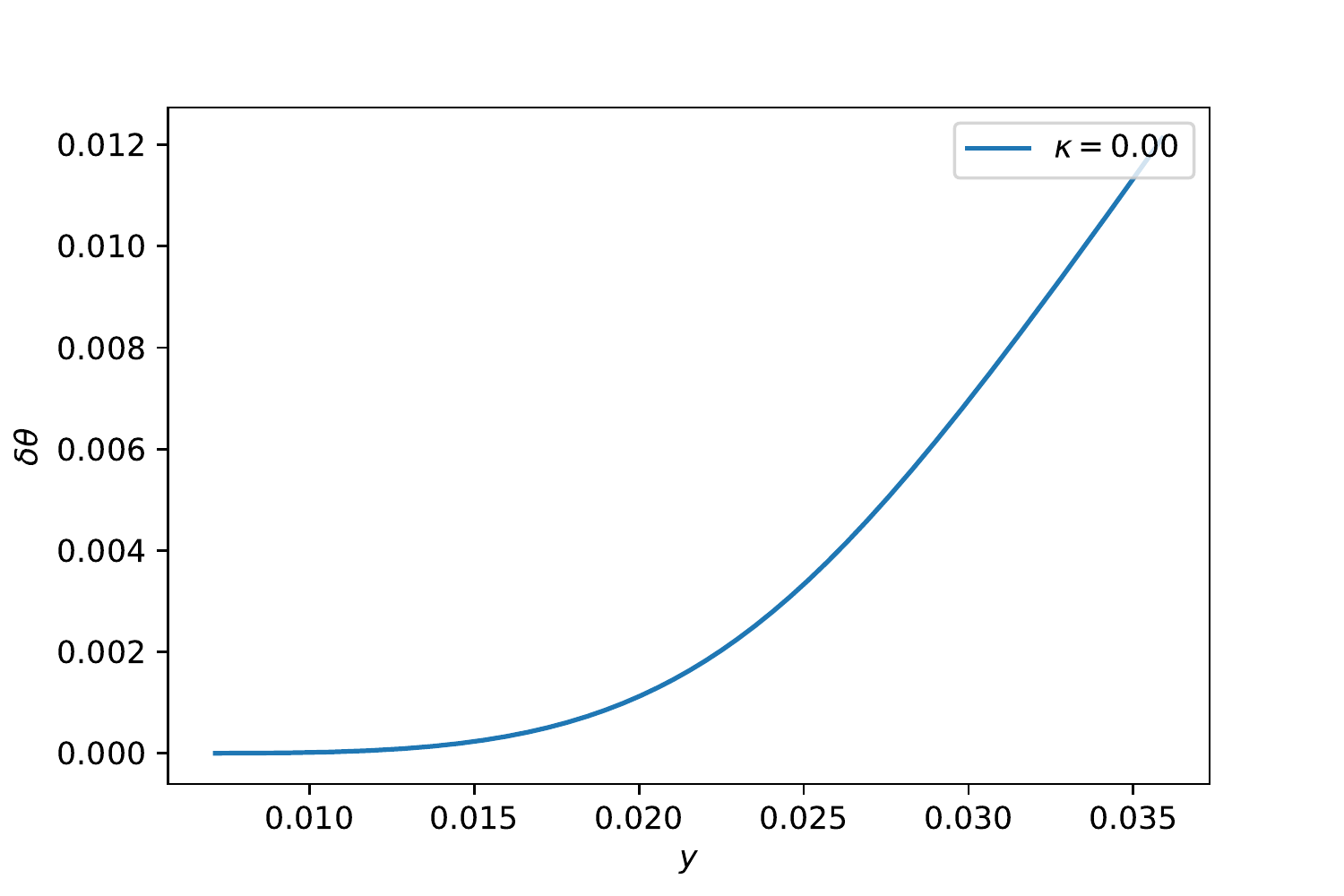}
	\caption{$\kappa = 0$ solution (hyperbolic case)}
	\label{fig:topBH_static_h_2}
	\end{minipage}
\end{figure}

\paragraph{Boundary behavior}
As go to the infinity, the metric function approaches $F_k\rightarrow1$ and 
$\partial F_k = 2ky - d\cdot r_\mathrm{m}^{d-2}y^{d-1}
\rightarrow0$ ($d=4$) for all cases ($\mathbb S$, $\mathbb H$ and $\mathbb T$).
Since this spacetime approaches flat AdS at infinity, the solution must approach the flat solution on the flat AdS space. 
The solution behaves at infinity as shown in \cite{Nagasaki:2011ue}.
It was $x_3 = \kappa y$, where $x_3$ is a usual Cartesian coordinate and $y$ is the radial direction.
In our coordinates it is $\kappa y = x_3 = X_3y = (r\cos\theta)y = \cos\theta$.
Expanding it around $\theta_0 = \pi/2$ gives
$\kappa y = \cos(\pi/2-\delta\theta)\sim\sin\delta\theta\sim\delta\theta$.
Then $\vartheta := \delta\theta\sim\kappa y$.
Then we have
$\vartheta'(\sigma)\sim\kappa y'(\sigma)$.
The gauge condition becomes 
$1 = y'^2 + \vartheta'^2F_k = (1+\kappa^2)y'^2$ at the AdS boundary.
By solving it for $y'$, $y(\sigma=0) = (1+\kappa^2)^{-1/2}$ and $\vartheta(\sigma=0) = \kappa(1+\kappa^2)^{-1/2}$. 
Therefore the boundary condition is 
\begin{equation}
y(\sigma=0) = 0,\quad
\vartheta(\sigma=0) = 0,\quad
y'(\sigma=0) = \frac1{\sqrt{1+\kappa^2}},\quad
\vartheta'(\sigma=0) = \frac{\kappa}{\sqrt{1+\kappa^2}}.
\end{equation}
The solution by the numerical calculation is shown in the following three figures:
Figure \ref{fig:topBH_static_s} shows the spherical case.
Figure \ref{fig:topBH_static_t} shows the toroidal case. 
As shown in Figure \ref{fig:topBH_static_h_1} and Figure \ref{fig:topBH_static_h_2} there is no solution which extends to the center of AdS spacetime for the hyperbolic case. 

\paragraph{No flux case}
When $\kappa=0$, the equations are simply
\begin{align}
y'' - y'\vartheta'S_k(\vartheta) + \frac4y(1-y'^2) - \frac12\vartheta'^2\partial F_k &= 0,\\
\vartheta'' + \vartheta'y'\frac{\partial F_k}{F_k} + \Big(\frac1{F_k} - \vartheta'^2\Big)S_k(\vartheta) 
 - \frac4yy'\vartheta' &= 0.
\end{align}
We defined 
\begin{equation}
S_k(\vartheta) := \frac{\partial}{\partial\theta}\log s_k(\theta)\Big|_{\theta=\pi/2-\vartheta}
= \begin{cases}
2\tan\vartheta& (\mathbb S)\\ 0& (\mathbb T)\\ -2\coth\vartheta& (\mathbb H)
\end{cases}.
\end{equation}
$(y,\vartheta) = (\sigma,\text{const})$ is a solution for the first equation and the second equation is satisfied as long as $S_k(\vartheta) = 0$.
This condition can be always satisfied only when $\vartheta = 0$ for $\mathbb S$.
For $\mathbb T$ case it is satisfied for any $\theta$ and for $\mathbb H$ there is no such a solution 
($|\coth x| > 1$ for all $x\in\mathbb R$).
We summarize the main result.
\begin{itemize}
\item The trivial solution $\theta = \mathrm{constant}$ can exist only for the spherical case and the toroidal and the hyperbolic case does not have such a solution. 
\end{itemize}

\section{Wilson loop operator}\label{sec:Wilson_loop}
We add a fundamental string which ends on the previous D5-brane.
We impose the Dirichlet boundary condition on the D5-brane and the AdS boundary.
We know that for spherical and toroidal cases the probe brane can extend into the black hole horizon.
We focus on these cases. 

We look for a static solution. 
The string is parametrized as
\begin{equation}
X^0(\tau,\lambda) = t(\tau),\;\;
X^1(\tau,\lambda) = y(\lambda),\;\;
X^2(\tau,\lambda) = \theta(\lambda),\;\;
\lambda\in(0,\lambda_1).
\end{equation}
We impose the Dirichlet boundary condition at the both ends of the string, $\lambda=0,\lambda_1$.
\paragraph{Boundary condition}
Suppose that the brane solution is given by the embedding $(Y(\sigma),\Theta(\sigma))$ and the string solution is given by $(y(\lambda),\theta(\lambda))$.
The  boundary conditions at both ends of the string are
\begin{subequations}\label{eq:staticBH_string_Dirichlet_bc}
\begin{align}
\lambda=0:&\quad
\theta(0) = \theta_0,\;\;
\theta'(\lambda)|_{\lambda=0} = 0,\\
\lambda=\lambda_1:&\quad
\theta(\lambda_1) = \Theta(\sigma_1),\;\;
y(\lambda_1) = Y(\sigma_1),\;\;
\frac{dY(\sigma)}{d\Theta(\sigma)}\cdot\frac{dy(\lambda)}{d\theta(\lambda)}\Big|_{y=y_1} = -F_k(y_1).
\end{align}
\end{subequations}

\paragraph{Trivial brane solution}

\begin{figure}[h]
\begin{center}
  \includegraphics[width = 12cm]{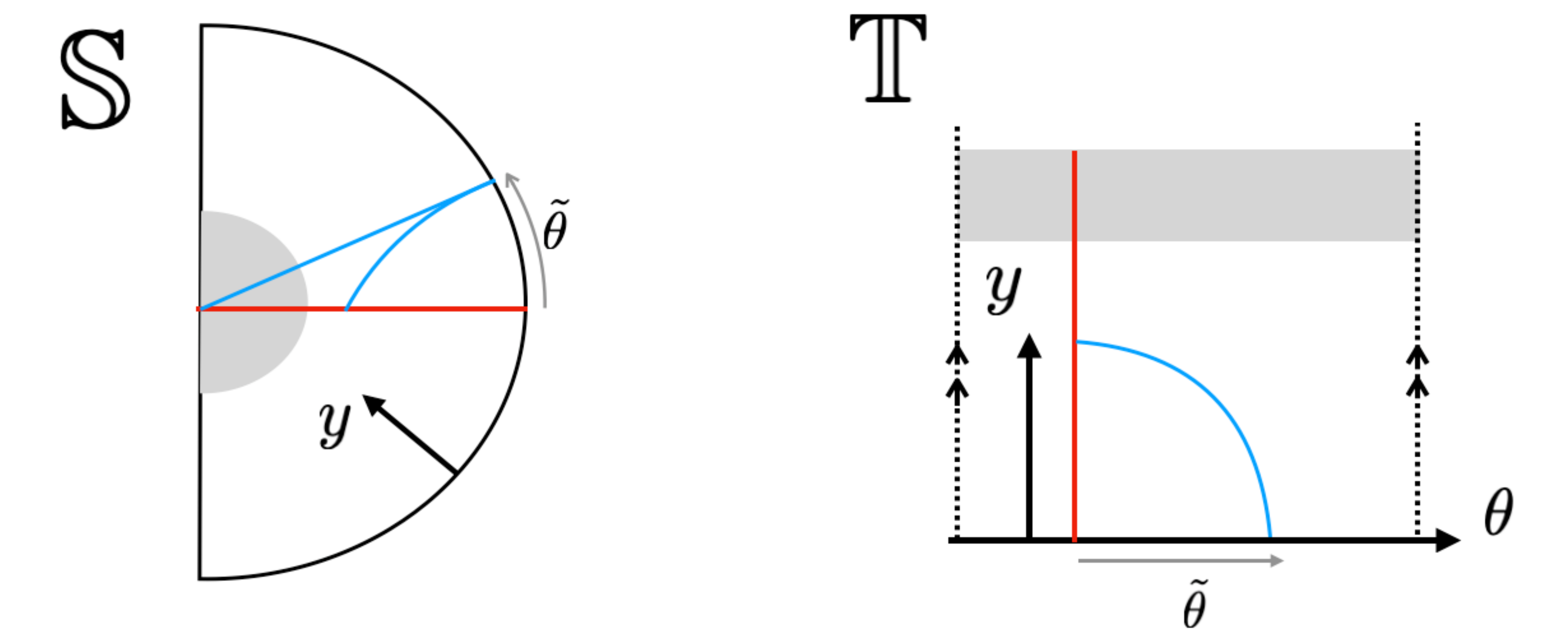}
  \caption{The probe brane and a fundamental string on the spherical and toroidal black hole spacetime}  
  \label{fig:D5_Fstring_on_ST}
\end{center}
\end{figure}

For the simple case, which has a trivial D5-brane solution and a fundamental string as shown Figure \ref{fig:D5_Fstring_on_ST}.
Since the form of the brane is given by $\Theta(\sigma) = \pi/2$, the boundary condition \eqref{eq:staticBH_string_Dirichlet_bc} is simplified as 
\begin{equation}
\theta(0) = \theta_0,\;\;
\theta(\lambda_1) = \frac\pi2,\;\;
\theta'(\lambda)|_{\lambda=0} = 0,\;\;
y'(\lambda)|_{\lambda = \lambda_1} = 0.
\end{equation}

We choose the conformal gauge $h_{ab} = \mathrm{diag}(1,1)$ and move to the Euclidean metric.
The Polyakov action of this fundamental string is 
\begin{align}
S_\mathrm{P} 
&= \frac1{4\pi\alpha'}\int d\tau d\lambda\sqrt h
 h^{ab}\partial_aX^\mu\partial_bX^\nu g_{\mu\nu}\nonumber\\
&= \frac1{4\pi\alpha'}\int d\tau d\lambda y^{-2}
 \Big(F_k(y)\dot t^2 + \frac{y'^2}{F_k(y)} + \theta'^2\Big),
\end{align}
where we used the metric  
\begin{subequations}
\begin{align}
ds^2 &= \frac1{y^2}\Big(F_k(y)dt^2 + \frac{dy^2}{F_k(y)}
  + d\Sigma_{k,3}^2\Big),\\
F_k(y) &= 1 + ky^2 - r_\mathrm{m}^2y^4.
\end{align}
\end{subequations}

The equations of motion for $t$ and $\theta$ are
\begin{equation}
\ddot t = 0,\;\;
\frac{d}{d\lambda}\big(y^{-2}\theta'\big) = 0.
\end{equation}
The first equation means $t$ is a linear function of $\tau$. 
The second can be solved by an integration constant $L$:
$\theta' = y^2L$.
The equation of motion for $y$ is 
\begin{align*}
 2y''\frac{y^{-2}}{F_k(y)} 
 + y'^2\frac\partial{\partial y}\Big(\frac{y^{-2}}{F_k(y)}\Big) 
 - \frac\partial{\partial y}\Big(y^{-2}F_k(y)\Big) + 2yL^2 = 0.
 \end{align*}
By multiplying $y'$ in both sides, it is summarized as a total derivative:
\begin{align*}
\frac{d}{d\sigma}(y'^2y^{-2}F_k(y)^{-1}
 - y^{-2}F_k(y) + y^2L^2) = 0.
\end{align*}
Therefore we find 
\begin{align}
y'^2y^{-2}F_k(y)^{-1} - y^{-2}F_k(y) + y^2L^2 &= \mathrm{const},\nonumber\\
y'^2 - F_k(y)^2 + y^4F_k(y)L^2 &= (\mathrm{const})\times y^2F_k(y).
\end{align}
The Virasoro constraint says the constant on the righthand side is zero. 

\paragraph{Virasoro constraint}
The Virasoro constraint: $T_{00} = -T_{11} = 0$.
\begin{equation}
T_{00} 
= \frac12(\partial_0X\cdot\partial_0X - \partial_1X\cdot\partial_1X).
\end{equation}
$T_{00} = 0$ gives 
\begin{equation}
y' = \sqrt{F_k(y)(F_k(y) - y^4L^2)}.
\end{equation}
By the boundary condition \eqref{eq:staticBH_string_Dirichlet_bc} at the AdS boundary, $y'(\lambda_1) = 0$, this $L$ is determined as
\begin{equation}
L^2 = F_k(y_1)y_1^{-4}.
\end{equation}
Then the integration constant $L$ is determined by the string position on the D5-brane (see Figure \ref{fig:simpleD5_fstr_ST}).
By substituting it, $y'$ is
\begin{equation}
y' = \sqrt{F_k(y)\Big(F_k(y) - \frac{y^4}{y_1^4}F_k(y_1)\Big)}.
\end{equation}
By this constraint the Polyakov action is simplified as
\begin{align}
\frac{dS_\mathrm{P}}{dt}
&= \frac1{4\pi\alpha'}\int_0^{y_1}d\sigma y^{-2}\Big[F_k(y) + \frac{y'^2}{F_k(y)} + \theta'^2\Big]\nonumber\\
&= \frac1{2\pi\alpha'}\int_0^{y_1}d\sigma y^{-2}F_k(y).
\end{align}

The integral for this action is diverge.
Hence we consider the two strings: one ends on the D5-brane and the other is on the black hole horizon.
We find the effect of the probe D5-brane by taking the difference of these two. 
When the fundamental string ends on the D5-brane
\begin{align}
S_\mathrm{P} 
&= \frac{T}{2\pi\alpha'}\int_0^{\sigma_1}d\sigma y^{-2}F_k(y)\nonumber\\
&= \frac{T}{2\pi\alpha'}\int_0^{y_1}\frac{dy}{y^2}\sqrt\frac{y^{-4}F_k(y)}{y^{-4}F_k(y) - y_1^{-4}F_k(y_1)}
\end{align}
On the other hand if the string go through the horizon, since in this case $L=0$, 
\begin{equation}
S_\mathrm{P0} = \frac{T}{2\pi\alpha'}\int_0^{y_\mathrm{h}}dyy^{-2}
= \frac{T}{2\pi\alpha'}\int_0^{y_1}\frac{dy}{y^2} 
 + \frac{T}{2\pi\alpha'}\Big[-y^{-1}\Big]_{y_1}^{y_\mathrm{h}},
\end{equation}
where the horizon is determined by $F_k(y_\mathrm{h}) = 0$.
Then their difference $\Delta S_\mathrm{P} := S_\mathrm{P} - S_\mathrm{P0}$ is 
\begin{equation}
\Big(\frac{T}{2\pi\alpha'}\Big)^{-1}\Delta S_\text{P}
= -\frac1{y_1} + \frac1{y_\mathrm{h}} 
 + \int_0^{y_1}\frac{dy}{y^2}\Bigg(-1 + \sqrt\frac{y^{-4}F_k(y)}{y^{-4}F_k(y) - y_1^{-4}F_k(y_1)}\Bigg).
\end{equation}


The variation of $\theta$ with respect to $y$ is 
\begin{equation}
\frac{d\theta}{dy} = \frac{\theta'}{y'}
= \sqrt\frac{y_1^{-4}F_k(y_1)}{F_k(y)(y^{-4}F_k(y) - y_1^{-4}F_k(y_1))}.
\end{equation}
The initial value of $\theta$ denotes as $\theta = \theta_1$. 
\begin{equation}
\frac\pi2 = \theta_1 + \int_0^{y_1}dy\sqrt\frac{y_1^{-4}F_k(y_1)}{F_k(y)(y^{-4}F_k(y) - y_1^{-4}F_k(y_1))}.
\end{equation}
We define $\tilde\theta_1(y_1)$ as the distance between the string the brane at the AdS boundary (Figure \ref{fig:D5_Fstring_on_ST}).
\begin{equation}
\tilde\theta_1(y_1) := \frac\pi2 - \theta_1 
= \int_0^{y_1}dy\sqrt\frac{y_1^{-4}F_k(y_1)}{F_k(y)(y^{-4}F_k(y) - y_1^{-4}F_k(y_1))}.
\end{equation}

\begin{center}
\begin{figure}[h]
	\begin{minipage}[t]{0.5\linewidth}
	\includegraphics[width=8cm]{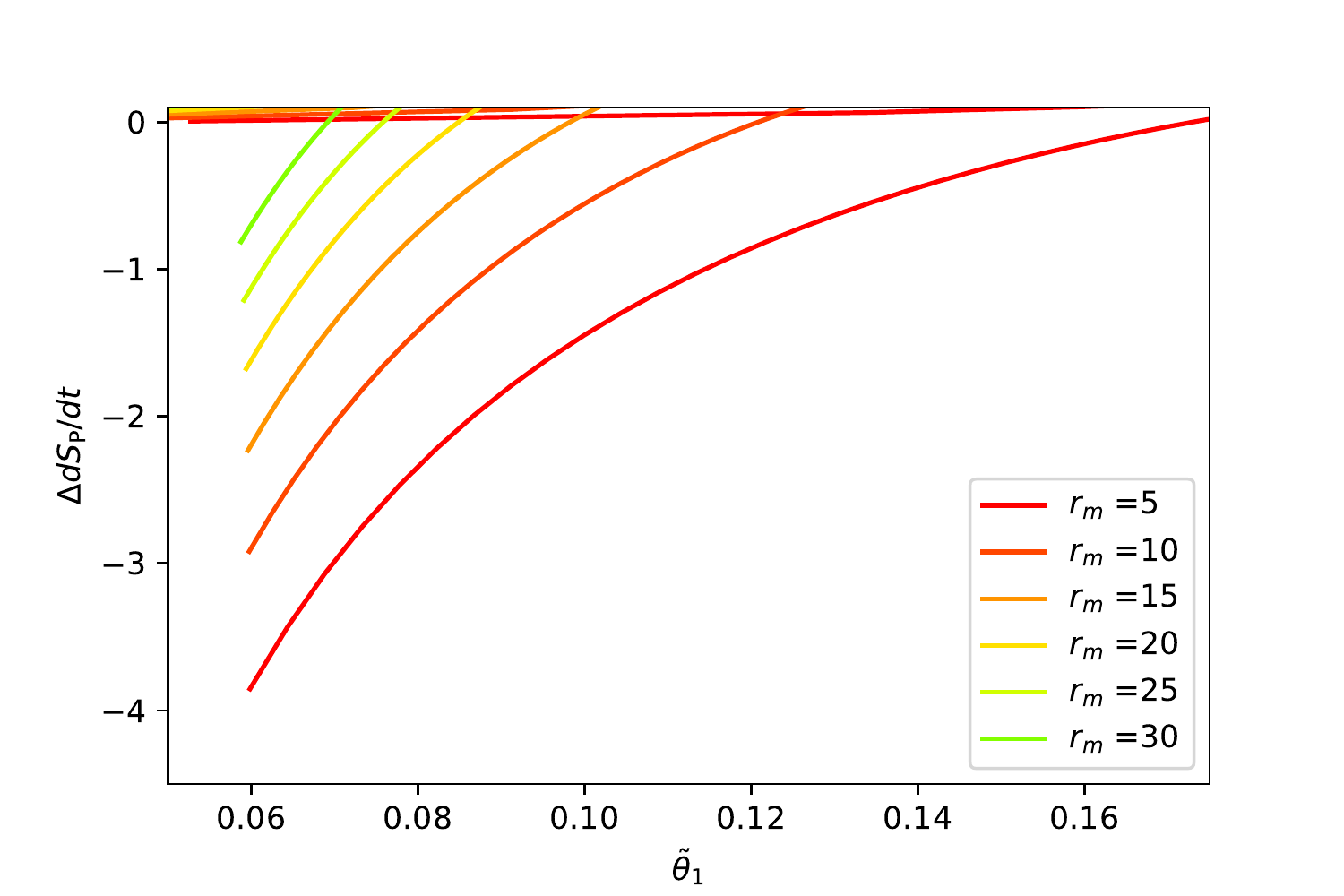}
	\caption{$\tilde\theta$ dependence of the action (sphere)}
	\label{fig:S_NGaction_tildetheta.pdf}
	\end{minipage}
\hspace{0.01\linewidth}
	\begin{minipage}[t]{0.5\linewidth}
	\includegraphics[width=8cm]{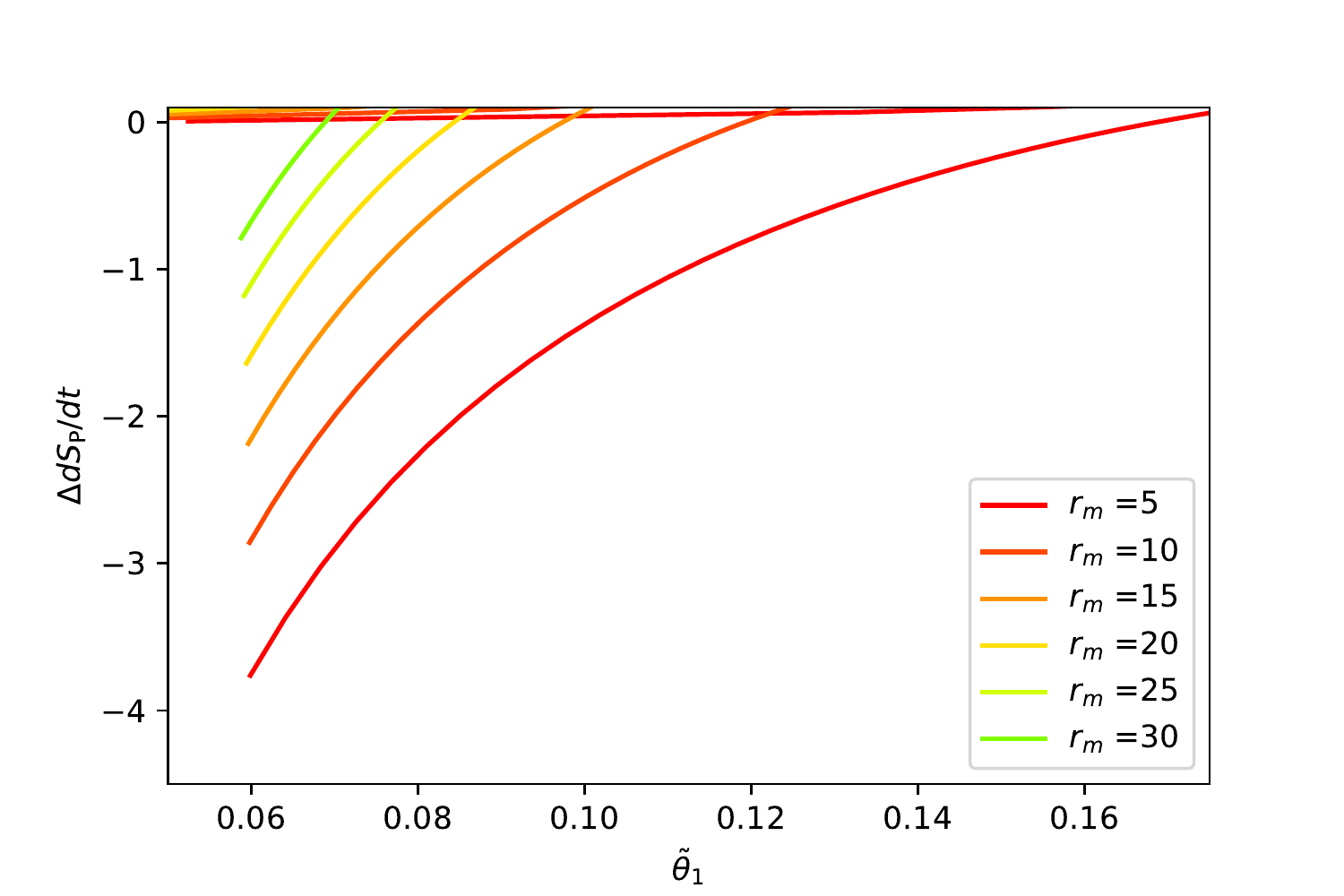}
	\caption{$\tilde\theta$ dependence of the action (torus)}
	\label{fig:T_NGaction_tildetheta.pdf}
	\end{minipage}
\end{figure}
\end{center}

The relation between the boundary position and the NG action is depicted in Figure \ref{fig:S_NGaction_tildetheta.pdf} and Figure \ref{fig:T_NGaction_tildetheta.pdf}.
Figure shows the comparing of all cases for mass $r_\text{m} = 3$.
This is a generalization we found in \cite{Nagasaki:2018dyz} to other curvatures $k=0,1$.

\newpage
\section{Nontrivial gauge flux}\label{sec:nontrivial_gaugeflux}
Here we would like to consider the case where the gauge flux is not zero: $\kappa\neq0$.
\begin{center}
\begin{figure}[h]
	\includegraphics[width=15cm]{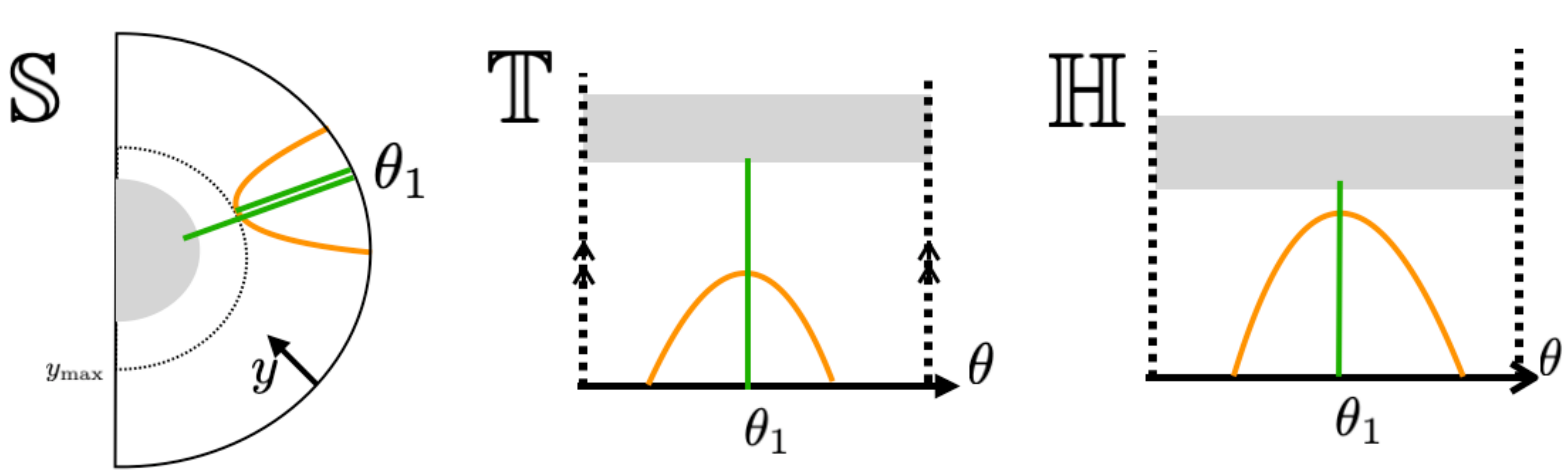}
	\caption{D5-brane with nontrivial gauge flux and strings}
	\label{fig:Lzero_strings.pdf}
\end{figure}
\end{center}

For $\kappa\neq0$, the brane has a nontrivial solution and we impose the boundary condition \ref{eq:staticBH_string_Dirichlet_bc}:
\begin{subequations}
\begin{align}
\sigma=0:&\quad
\theta(0) = \theta_0,\;\;
\theta'(\lambda)|_{\lambda=0} = 0,\\
\sigma=\sigma_1:&\quad
\theta(\lambda_1) = \Theta(\sigma_1),\;\;
y(\lambda_1) = Y(\sigma_1),\;\;
\frac{dY(\sigma)}{d\Theta(\sigma)}\cdot\frac{dy(\lambda)}{d\theta(\lambda)}\Big|_{y=y_1} = -F_k(y_1).
\end{align}
\end{subequations}

The equations of motion for the D5-brane are 
\begin{align}
Y'' + Y'\Theta'\partial\log s_k
 + \frac4y(1-Y'^2-\tilde\kappa\Theta')
 - \frac12\Theta'^2\partial F_k &= 0,\\
\Theta''F_k + \Theta'Y'\partial F_k
 + (\Theta'^2F_k - 1)\partial\log s_k
 - \frac4y(Y'\Theta'F_k - \tilde\kappa Y') &= 0.
\end{align}
These differential equations are written as first-differential equations as
\begin{align}
\frac{dY}{d\sigma} &= P_Y,\quad
\frac{d\Theta}{d\sigma} = P_\Theta,\label{eq:diffeq_YTheta_ytheta_1}\\
\frac{dP_Y}{d\sigma} 
&= -P_YP_\Theta\partial\log s_k
 - \frac4Y(1-P_Y^2-\tilde\kappa P_\Theta)
 + \frac12P_\Theta^2\partial F_k,\\
\frac{dP_\Theta}{d\sigma}
&= -P_YP_\Theta\frac{\partial F_k}{F_k}
  + \frac4Y\Big(P_YP_\Theta + \tilde\kappa\frac{P_Y}{F_k}\Big)
  + \Big(P_\Theta^2 - \frac1{F_k}\Big)\partial\log s_k.
\end{align}
The string equation of motion is
\begin{equation}
y' = \sqrt{F_k(y)\Big(F_k(y) - \frac{y^4}{y_1^4}F_k(y_1)\Big)}.
\end{equation}

We consider a special solution where $L=0$.
We denote the value of $y$ where $p_y=0$ is satisfied as $y_\mathrm{max}$ (see Figure \ref{fig:Lzero_strings.pdf}). 
In this case, the action is 
\begin{align}
\Delta S_\mathrm{P} 
:= S_\mathrm{P} - S_\mathrm{P0}
= \frac{T}{2\pi\alpha'}\Big(
 \int_0^{y_\mathrm{h}}\frac{dy}{y^2} - \int_0^{y_\mathrm{max}}\frac{dy}{y^2}
 \Big)
= \frac{T}{2\pi\alpha'}\Big(
 \frac1{y_\mathrm{max}} - \frac1{y_\mathrm{h}}
 \Big).
\end{align}
In the first term $y_\text{max}$ depends on the gauge flux $\kappa$ and mass, while the second term only depends on mass.

The boundary conditions on the both sides were
\begin{subequations}
\begin{align}
\sigma=0:&&
\theta(0) &= \theta_0,&
\theta'(\lambda)|_{\lambda=0} &= 0,\\
\sigma=\sigma_1:&&
\theta(\sigma_1) &= S(y(\sigma_1)),&
\Big(\frac1{F_k(y)}Y'(\sigma)y'(\lambda) + \Theta'(\sigma)\theta'(\lambda)\Big)\Big|_{y=y_1} &= 0.
\end{align}
\end{subequations}
Now $\theta'(\lambda) = 0$ for arbitrary $\lambda$ and $Y'(\sigma) = 0$ at the maximum point of $Y$. Then the second and last conditions are satisfied.

\begin{center}
\begin{figure}[h]
	\begin{minipage}[t]{0.5\linewidth}
	\includegraphics[width=8cm]{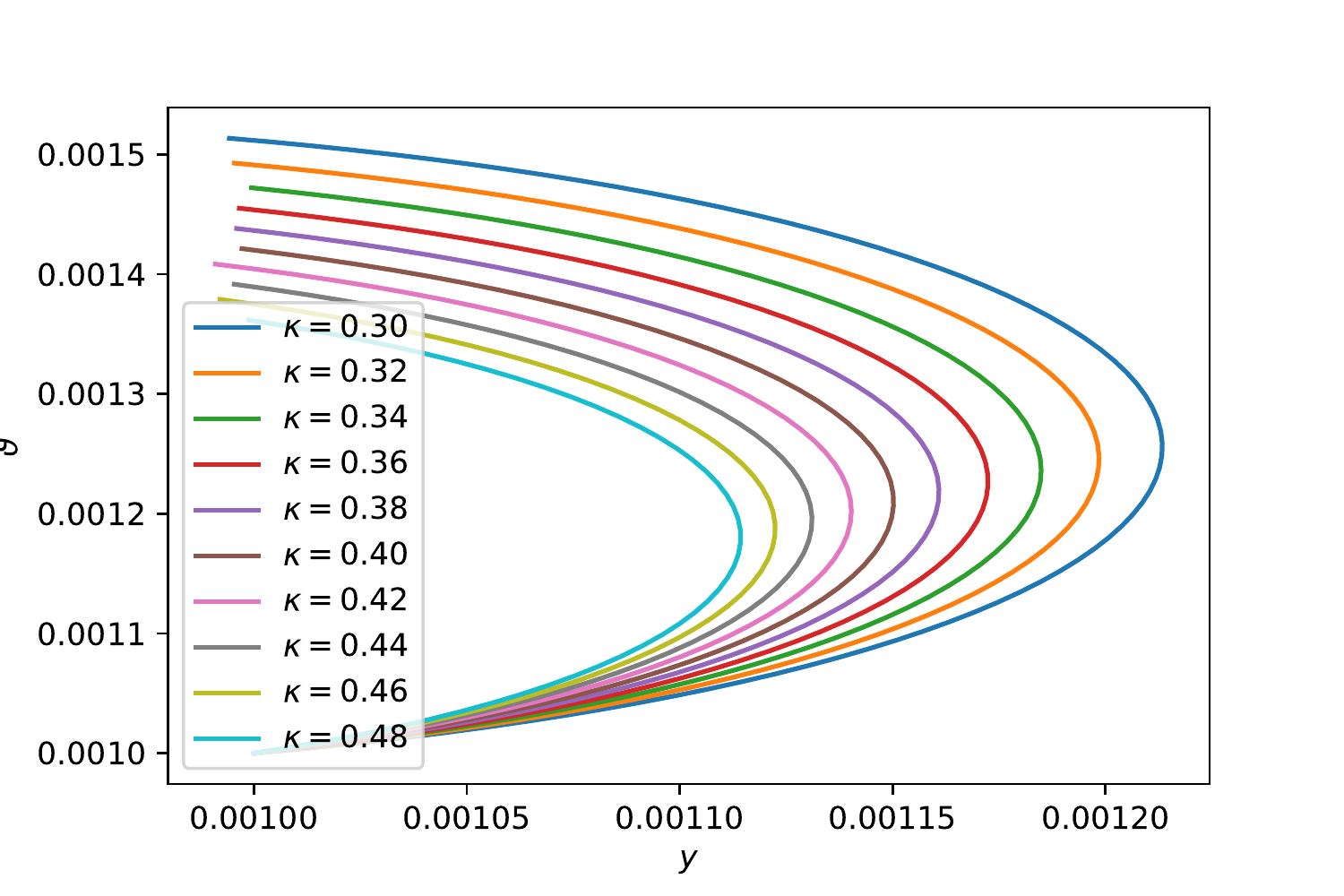}
	\caption{Flux dependence for fixed mass (spherical case)} 
	\label{fig:D5_ADSBH_rm10_y_vartheta_S}
	\end{minipage}
\hspace{0.01\linewidth}
	\begin{minipage}[t]{0.5\linewidth}
	\includegraphics[width=8cm]{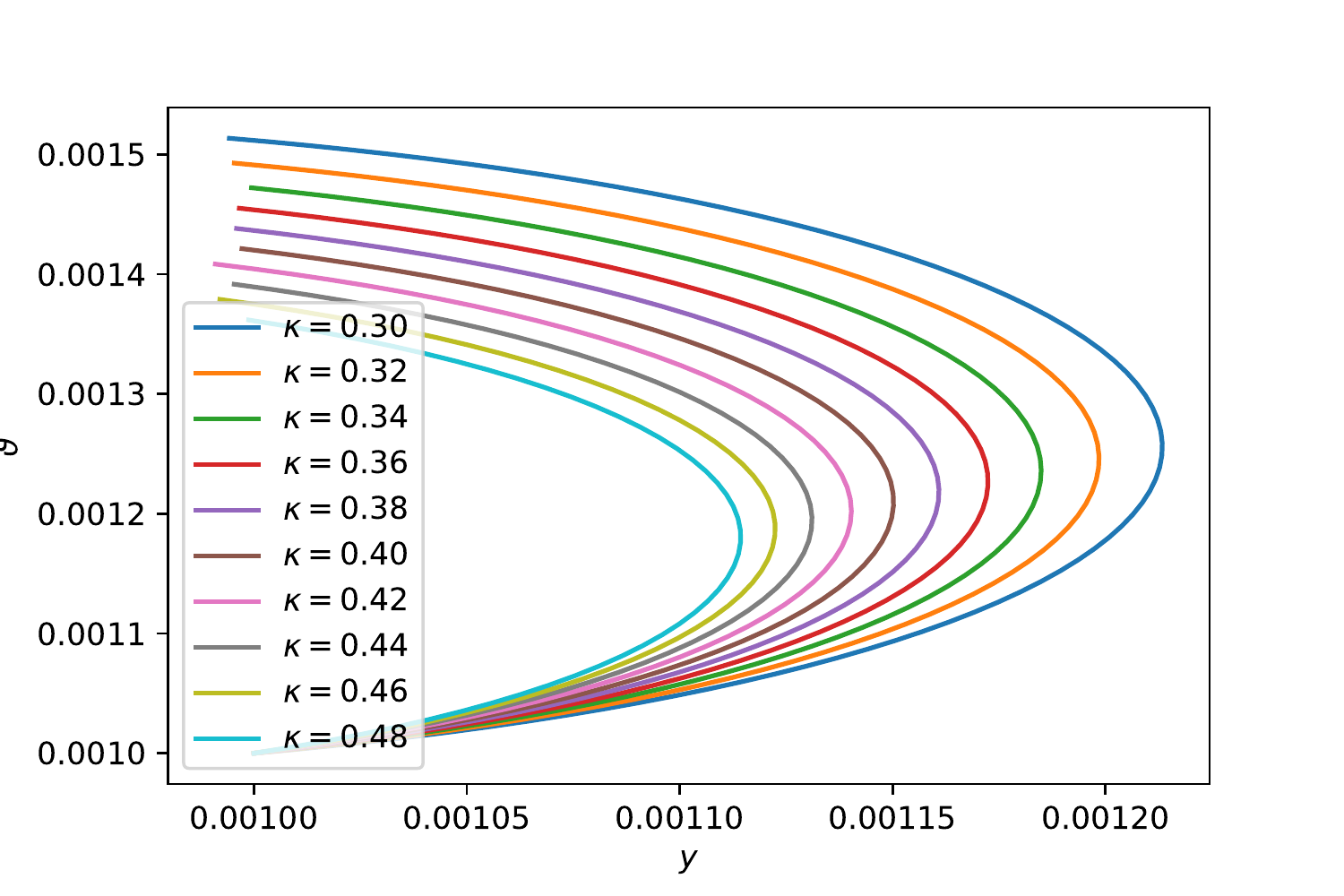}
	\caption{Flux dependence for fixed mass (toroidal case)}
	\label{fig:D5_ADSBH_rm10_y_vartheta_T}
	\end{minipage}
\hspace{0.01\linewidth}
	\begin{minipage}[t]{0.5\linewidth}
	\includegraphics[width=8cm]{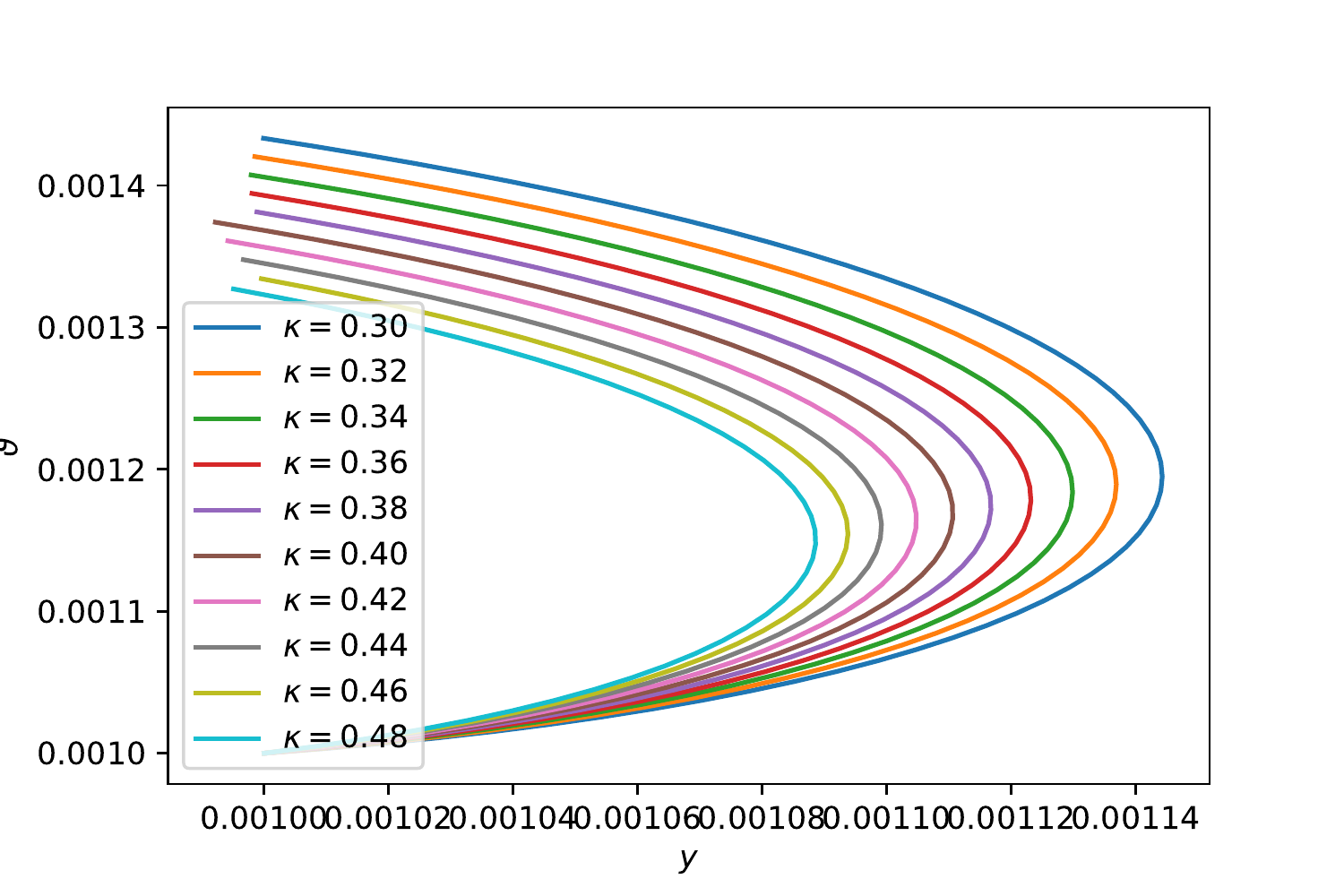}
	\caption{Flux dependence for fixed mass (hyperbolic case)}
	\label{fig:D5_ADSBH_rm10_y_vartheta_H}
	\end{minipage}
\end{figure}
\end{center}

The horizon $y_\mathrm{h}$ is determined by 
\begin{equation}
r_\mathrm{m}^2y_\mathrm{h}^4 - ky_\mathrm{h}^2 - 1 = 0,\;\;
y_\text{h} = \bigg(\frac{k+\sqrt{4r_\mathrm{m}^2 + k^2}}{2r_\mathrm{m}^2}\bigg)^{1/2}.
\end{equation}
We only need to find the value $y_\text{max}$ to calculate the action. 
$y_\text{max}$ defined in \eqref{eq:diffeq_YTheta_ytheta_1}  is the value where $P_Y$ becomes zero. 
It is plotted in Figure \ref{fig:D5_ADSBH_rm10_y_vartheta_S}, Figure \ref{fig:D5_ADSBH_rm10_y_vartheta_T} and Figure \ref{fig:D5_ADSBH_rm10_y_vartheta_H}.

\begin{center}
\begin{figure}[t]
	\begin{minipage}[t]{0.5\linewidth}
	\includegraphics[width=8cm]{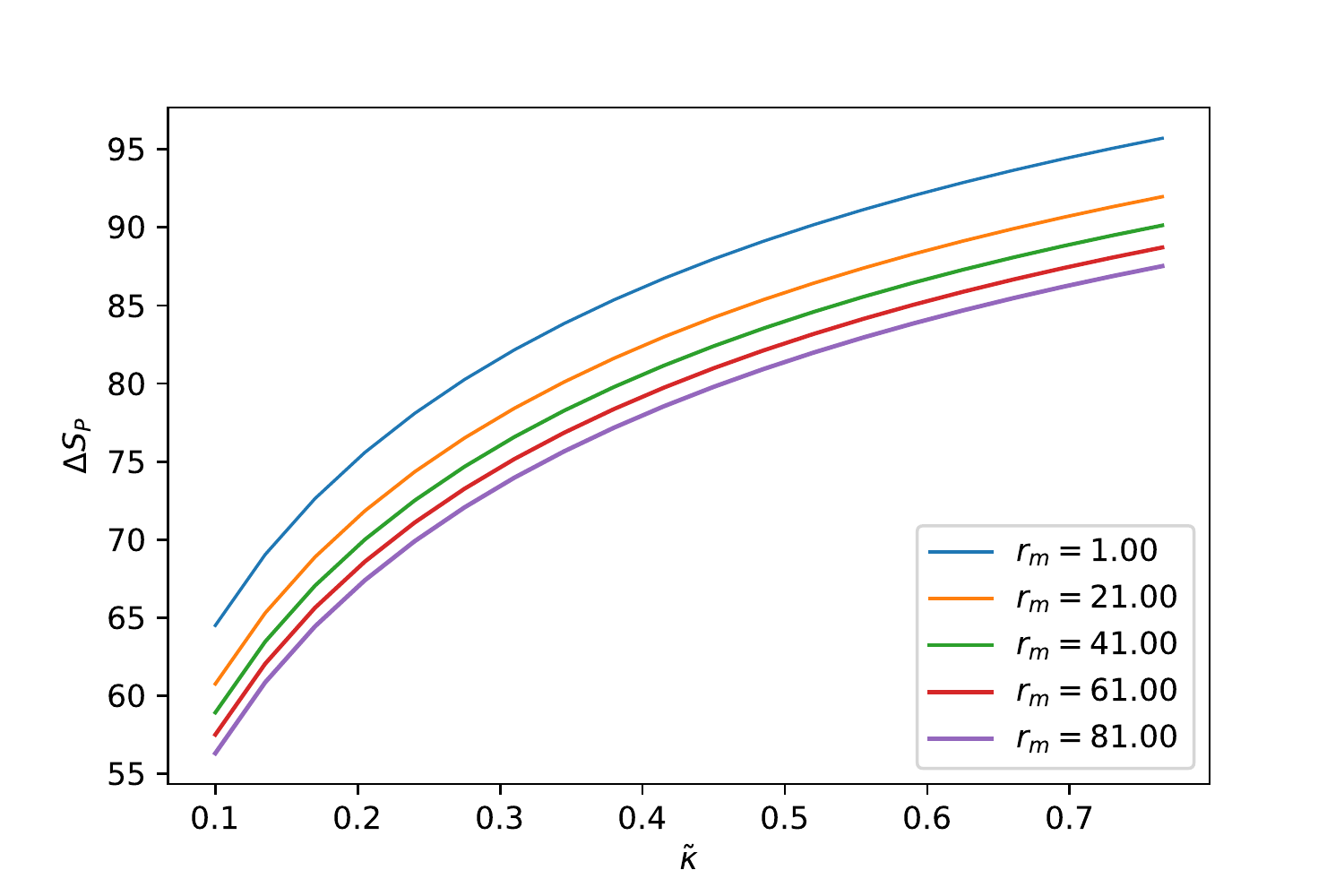}
	\caption{Action (spherical case)}
	\label{fig:D5_ADSBH_Action_flux_S}
	\end{minipage}
\hspace{0.01\linewidth}
	\begin{minipage}[t]{0.5\linewidth}
	\includegraphics[width=8cm]{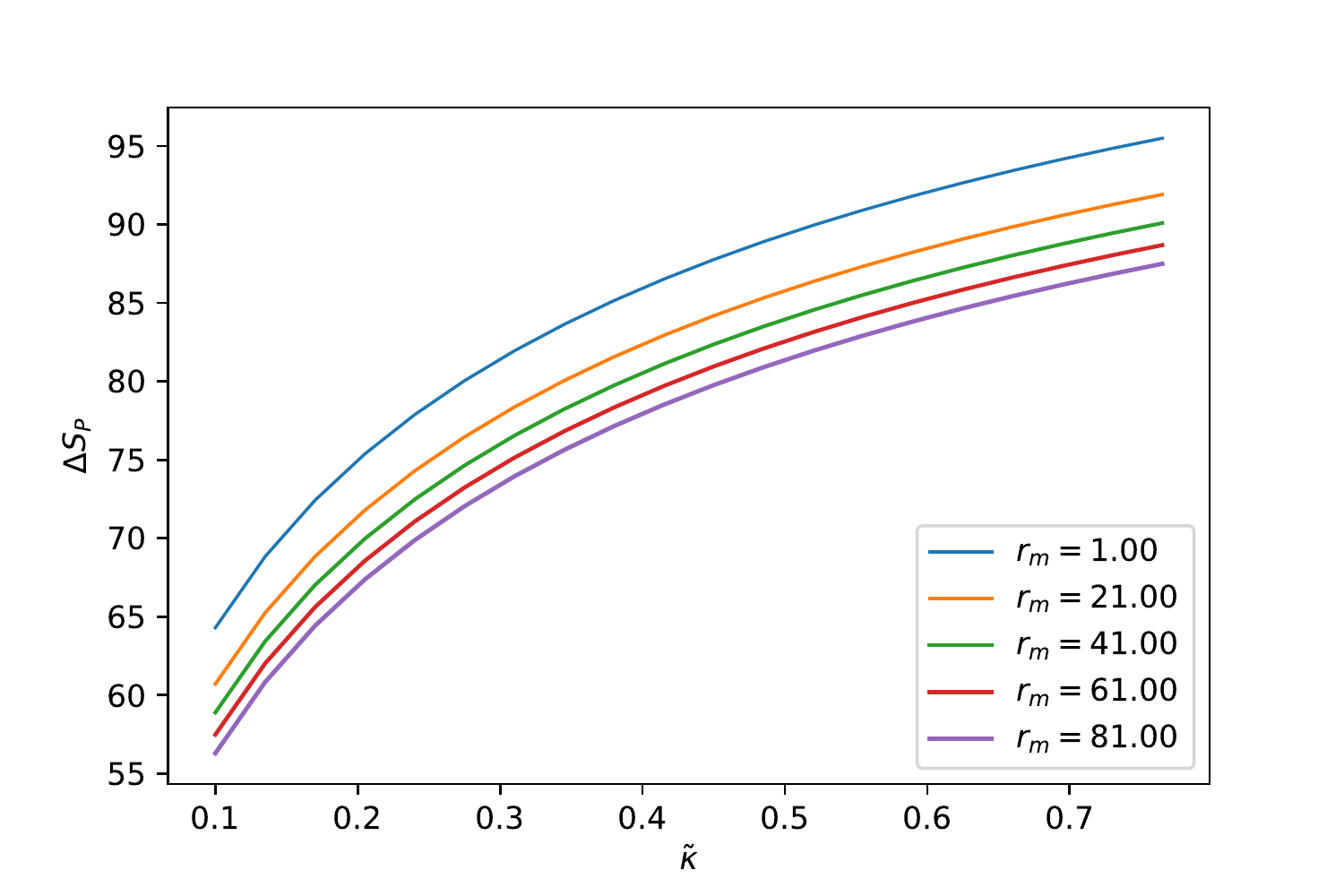}
	\caption{Action (toroidal case)}
	\label{fig:D5_ADSBH_Action_flux_T}
	\end{minipage}
\hspace{0.01\linewidth}
	\begin{minipage}[t]{0.5\linewidth}
	\includegraphics[width=8cm]{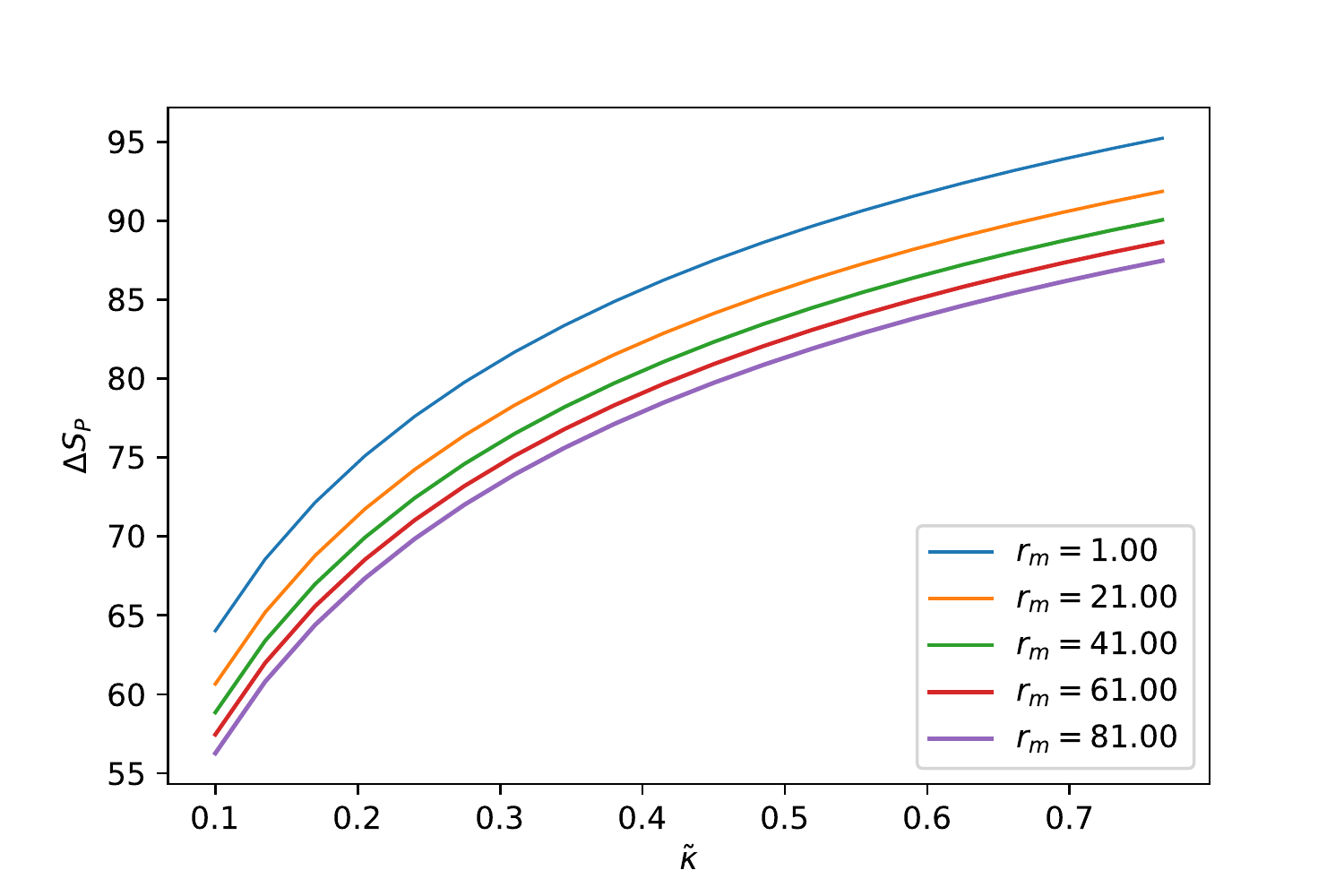}
	\caption{Action (hyperbolic case)}
	\label{fig:D5_ADSBH_Action_flux_H}
	\end{minipage}
\end{figure}
\end{center}

\newpage
\section{Discussion}
The interface is a nonlocal object which determines the boundary condition between two gauge theories.
This operator corresponds to a probe D5-brane in the bulk AdS spacetime.
In this paper we studied the behavior of this solution in the three different black hole spacetimes: sphere, toros and hyperbolic space.
Unlike the flat AdS spacetime \cite{Nagasaki:2011ue} this D5-brane has a nontrivial configuration.
We found that for black hole spacetimes which has spherical and toroidal structures the probe D5-brane can exist inside the horizon only when the gauge flux on D5-brane is zero and bend so that it avoids the horizon if the brane has a nonzero gauge flux.
For hyperbolic space there is no such a solution even if the gauge flux is zero.

From the perspective of gauge theory two different gauge groups $\mathrm{SU}(N)$ and $\mathrm{SU}(N-k)$ can exist on the both sides of the interface, where the difference of the gauge group $k$ is related to the gauge flux on the D5-brane \cite{Nagasaki:2011ue} as
\begin{equation}\label{eq:kappa_k_relation}
k = -\frac{T_5}{T_3}\int\mathcal F = \frac\kappa{\pi\alpha'}.
\end{equation}
Here $T_5$ and $T_3$ are tension of the D5-brane and the D5-brane.
We introduced the gauge flux $\kappa$ on the D5-brane. 

By adding a fundamental string we calculate the potential energy between the interface and a Wilson loop.
For a trivial D5-brane solution ($\kappa=0$) we find the behavior of potential energy between the interface and the test particle (Figure \ref{fig:S_NGaction_tildetheta.pdf} for sphere and Figure \ref{fig:T_NGaction_tildetheta.pdf} for torus).

We were interested in a more general case, where the probe brane has non zero gauge flux.
In this case where gauge theories with different gauge groups exist on the both sides of the interface.
For a special case (Figure \ref{fig:Lzero_strings.pdf}) where the string has the trivial configuration, we found the potential energy by the Nambu-Goto action (Figure \ref{fig:D5_ADSBH_Action_flux_S}, Figure \ref{fig:D5_ADSBH_Action_flux_T} and Figure \ref{fig:D5_ADSBH_Action_flux_H}).
For non trivial gauge flux, we found this potential energy has the similar behavior although the behavior of zero flux solution is very different for the hyperbolic spacetime case: this case has no solution which approaches the black hole horizon (Figure \ref{fig:topBH_static_h_2}).
For all topologies (sphere, torus and hyperbolic space) the potential energy between the interface and the Wilson loop becomes an increasing function of $\kappa$ and also increasing with respect to the mass of black holes. 

An important direction is to generalize the correspondence to a time-dependent process by considering  a drag force \cite{Gubser:2006bz}, which was described by a string with constant velocity, or considering the rotating black holes.
\\

\section*{Acknnowledgments}
This research of is supported by Department of Physics, Toho University.

\providecommand{\href}[2]{#2}\begingroup\raggedright\endgroup

\end{document}